\documentstyle[12pt]{article}
\newcommand{\blankline}{\vskip .3cm}
\newcommand{\f}{\begin{equation}}
\newcommand{\ff}{\end{equation}}
\newcommand{\be}{\begin{equation}}
\newcommand{\ee}{\end{equation}}
\newcommand{\bea}{\begin{eqnarray}}
\newcommand{\eea}{\end{eqnarray}}
\begin{document}
\centerline{\LARGE String theories with deformed energy momentum 
relations,} \centerline{\LARGE and a  possible
non-tachyonic bosonic string} \blankline \blankline \rm
\centerline{Jo\~ao Magueijo${}^\#$ and Lee Smolin${}^{*}$}
\blankline \centerline{\it ${}^\#$The Blackett Laboratory,Imperial
College of Science, Technology and Medicine } \centerline{\it
South Kensington, London SW7 2BZ, UK} \centerline{\it
${}^*$Perimeter Institute for Theoretical Physics} \centerline{\it
Waterloo, Canada  N2J 2W9} \centerline{\it  \ \ and }
\centerline{\it Department of Phyiscs, University of Waterloo}

\blankline \blankline \blankline \blankline \centerline{January,
2004} \blankline \blankline \blankline \blankline

\centerline{ABSTRACT}
We consider a prescription for introducing deformed dispersion relations
in the bosonic string action. We find that in a subset of such theories
it remains true that the embedding coordinates propagate linearly
on the worldsheet.   While both the string modes and
the center of mass propagate with deformed dispersion relations, the
speed of light remains energy independent.
We  consider the
canonical quantization of
these strings, and find that it is possible to choose theories so that
ghost modes still decouple, as usual.
 We also find that there are examples where the
tachyon is eliminated from the spectrum of the free bosonic string.
\vfill
\blankline

\blankline
lsmolin@perimeterinstitute.ca,\\    j.magueijo@imperial.ac.uk
\eject
\tableofcontents

\vfill

\section{Introduction}

Several experimental and theoretical developments point to the
possibility that the usual relation between energy and momentum
valid within the special theory of relativity,
\f
E^2 = p^2 + m^2
\label{usual}
\ff
may be modified at Planck scales.  For instance, the  high-energy cosmic
ray anomalies~\cite{review,crexp} may be solved if there are Planck
scale departures from these
relations~\cite{amel,amel1,liouv,cosmicray,leejoao1}.
It has also been shown that one may establish an observer independent
border between the classical and quantum pictures of
space-time~\cite{amelstat,gli,leejoao}, by means of specially designed
deformed dispersion relations.

The usual argument linking these deformations to quantum gravity is
simple~\cite{leejoao}. The combination of gravity ($G$), the quantum ($\hbar$)
and relativity ($c$) gives rise to the Planck length,
\f
L_P = \sqrt{\hbar G \over c^3 }
\ff
or its inverse, the Planck energy
\f
E_P = \sqrt{\hbar  c^5 \over G} .
\ff
These scales mark thresholds beyond which the old description of
spacetime breaks down and qualitatively new phenomena are expected to
appear.  However, the new theory has to agree with special
relativity for experiments probing the nature of space-time at energy scales
much smaller than $E_P$. The question, then, arises:  {\it in whose reference
frame are $L_P$ and $E_P$ the thresholds for new phenomena?}
It is clear that the Lorentz-Fitzgerald contraction cannot apply all
the way down to the Planck scale, if we are to avoid introducing
preferred frame in quantum gravity. A possible way out is the introduction
of non-linear realizations of the Lorentz group, associated with
deformed dispersion relations. This possibility, sometimes called,
``doubly special relativity" has been explored in a number of
recent papers\cite{amelstat}-\cite{leejoao}.

The most important reason to investigate this possibility is that
it is an hypothesis that will be testable in experiments to be
carried out over the next decade. What if a combination of
low energy and astrophysical experiments found evidence of
this kind of theory, such as energy dependent speed of light and/or
modifications in relativistic energy-momentum conservation laws?
What would be the implications for various candidate quantum theories
of gravity? Given that the relevant experiments are planned, it is
important if the different approaches to quantum
gravity make predictions for their outcomes.

It has recently been shown that this possibility is realized in
the case of quantum gravity coupled to point particles in $2+1$
dimensions\cite{2+1}.  This shows that there are physically
sensible interacting quantum theories of gravity that are
described in terms of deformed dispersion relations. As far as
$3+1$ dimensions is concerned, there are as yet no definitive
results, but there are preliminary
indications\cite{loopdef,loopdef1} that such deformations may
emerge naturally in loop quantum gravity~\cite{rovelli}.

In this paper we thus take up the question of whether phenomena
associated with deformed dispersion relations may be realized in a
consistent string theory~\cite{pol,GSW,stringref}.  For simplicity,
we study here only the bosonic string.

In Section~\ref{eqns} we present the formalism for introducing
deformed dispersion relations into the action of the bosonic
string. We find that in most cases
the vibrational modes do not  linearize, however there is a large
class of examples in which they do.
We study these in  Section~\ref{slns} and we are able then to
first quantize the theory.
Next, we examine the constraints imposed at the quantum level.
In Section~\ref{quantum} we  study the Virasoro algebra and spectrum
of the deformed strings.  

In the following section (Section~\ref{examples}) we look at three
simple examples. The first employs a simple deformation of the
energy-momentum relations, studied for particles in
\cite{leejoao}, in which there is a maximum invariant mass (or
rest energy to be more precise). We show that this remains the
case in string theory, so that the rest energies of the string
excitations are bounded from above.

In the next two examples, the deformed energy-momentum relations are
chosen so that the tachyonic mode of the free bosonic string is eliminated.
We find this an intriguing result, as the possibility of a consistent,
tachyon free
bosonic string would obviously be of great importance for
string theory.

\section{The action and equations of motion}\label{eqns}

It is possible to write the bosonic string action in the form:
\begin{equation}
S=\int d\sigma d\tau (\dot x^a p_a -N {\cal H} - M {\cal D})
\end{equation}
where, $N$ and $M$ are Lagrange multipliers, ${\cal H} $ is the
Hamiltonian constraint, and ${\cal D}$ is the constraint associated with
spatial diffeomorphism invariance on the string.   Together they generate 
diffeomorphisms on the worldsheet.   However we now write the
Hamiltonian constraint as
\begin{equation}\label{ham}
{\cal H}= {f\over 2T}\eta ^{ab}p_a p_b+ {Tg\over 2}\eta _{ab}
x^{a'} x^{b'}
\end{equation}
where $T$ is the string tension and the dash represents a
derivative with respect to $\sigma$. It can be checked that
this action is reparameterization invariant. Here $f$ and $g$ are
functions of the total energy
\begin{equation}
{\bf P}_0=\int d\sigma p_0
\end{equation}
and are expected to encode all deviations from linear special
relativity. Furthermore
\begin{equation}
{\cal D}={\sqrt {fg}} p^a x'_a
\end{equation}
With these modifications we can still write the algebra of
constraints as
\begin{equation}\label{vir}
L_n=\int d\sigma e^{in\sigma} 2({\cal H}+{\cal D})= \int d\sigma
e^{in\sigma} {\left({\sqrt{f\over T}}p_a +{\sqrt
{Tg}}x'_a\right)}^2
\end{equation}

Notice that the theory we have just proposed is not a mere redefinition
of the string tension. Firstly $f$ and $g$ multiply $T$ differently
in the various terms of the constraints. Secondly, even if one considers
each of these terms on its own, it looks as if the tension is being
renormalized by a factor dependent on the total string energy. The latter
is the zero component of a vector. Thus such a ``renormalized''
tension would no longer be a scalar, invariant for all observers.
One may expect a variety of new phenomena/pathologies to appear,
an expectation  we shall soon confirm.

The equations of motion can be found, e.g. from Hamilton's
equation, and we note that the space and time coordinates in
target space are now to be treated differently. For the spacial
coordinates we find:
\begin{eqnarray}
\dot x^i&=&{\partial{\cal H}\over \partial p_i}={f\over T}p^i\\
\dot p_i&=&- {\partial{\cal H}\over \partial
x^i}=Tg\partial_\sigma^2 x_i
\end{eqnarray}
from which we may infer
\begin{equation}
\ddot x^i-fg\partial^2_\sigma x^{i}=0
\end{equation}
Hence we find our first result: the $x^i$ coordinates satisfy the
wave equation, but the speed of light on the string depends upon
the total energy ${\bf P}_0$ stored in the string:
\begin{equation}
c_s=\sqrt {fg}
\end{equation}
We denote this speed as $c_s$ because it is really a speed of
sound, i.e. a speed of propagation of vibrations along the string.
It can be easily proved that ${\bf P}_0$ remains a constant of
motion for general $f$ and $g$.

A similar set of equations may be found for the time target space
coordinate
\begin{eqnarray}
\dot x^0&=&{\partial{\cal H}\over \partial p_i}={f\over T}p^0 +{f'\over
2T}p^2
+ {Tg'\over 2} {x'}^2\\
\dot p_0&=& -{\partial{\cal H}\over \partial
x^0}=Tg\partial_\sigma^2 x_0
\end{eqnarray}
As a result we find that the $x^0$ coordinate in general satisfies
a rather complicated non-linear equation, coupled to the $x^i$
coordinates. The only exception occurs if $f=g$. Then the new
terms are proportional to  ${\cal H}$ and so vanish as a result of
the Hamiltonian constraint. Hence we learn that the deformations
with $f=g$ play a special role in string theory, as they preserve
the equation
\begin{equation}\label{wave}
\ddot x^a-fg\partial^2_\sigma x^{a}=0
\end{equation}
for $a=0,i$.

It is very difficult to find solutions in the coupled case. The
bosonic string has become non-linear and the quanta travelling
along the string interact with each other. This unpleasant property
already plagues standard $p$-branes with $p>1$.

In order to build further intuition about the meaning of the case
$f=g$, in the Appendix we consider the analogue construction for
point particles.

\section{Solutions and canonical quantization}\label{slns}
The particular case where the string remains linear is therefore
the only one where we shall be able to perform a full study.  In
this case we may introduce light-cone variables:
\begin{equation}
\sigma ^\pm=\sqrt {fg}\tau\pm\sigma
\end{equation}
It is important that in this definition the speed of sound
multiples the coordinate $\tau$, rather than divide the
coordinate $\sigma$. In terms of $\sigma^\pm$
the wave equation (\ref{wave}) becomes
\begin{equation}
\partial_+\partial_-x^a=0
\end{equation}
Its most general solution is:
\begin{equation}
x^a={\bf x}^a+{{\bf P}^a_{CM}\over \pi T} \tau +
v(\sigma^+)+w(\sigma^-)
\end{equation}
where ${\bf x}^a$ and ${\bf P}_a^{CM}$ are center of mass
integration constants. If we consider an open string with
$\sigma\in [0,\pi]$ we can therefore write:
\begin{equation}\label{xs}
x^a={\bf x}^a+{{\bf P}^a_{CM}\over \pi T} \tau +{i\over {\sqrt {\pi T}}}
\sum_{n\neq 0} {\alpha^a_n\over n} e^{-in{\sqrt{fg}}\tau} \cos(
n\sigma)
\end{equation}
Note that we must have $\alpha^a_{-n}=\alpha^{a\dagger}_n$, for
$x^a$ to be real. The associated string momentum is:
\begin{equation}\label{ps}
p_a={T\over f}\dot x_a={{\bf P}^a_{CM}\over \pi f}+ \sqrt {gT\over
\pi f}\sum_{n\neq 0}{\alpha^a_n} e^{-in{\sqrt{fg}}\tau} \cos(
n\sigma)
\end{equation}

We now proceed to canonically quantize this string, starting
from the equal-time commutation relations:
\begin{equation}
[x^a(\sigma,\tau),p_b(\sigma',\tau)]=i\delta(\sigma-\sigma')\delta^a_b
\end{equation}
Using
\begin{equation}
\delta(\sigma-\sigma')={1\over \pi}{\left(1+2\sum_{n=1}^\infty
\cos(n\sigma)\cos(n\sigma')\right)}
\end{equation}
we thus arrive at:
\begin{eqnarray}
{\left[{\bf x}^a,{\bf P}^{CM}_b\right]}&=&if\delta^a_b\\
{\left[\alpha_n^a,\alpha_{-n}^b\right]}&=&\sqrt{f\over
g}n\eta^{ab}\label{alphcom}
\end{eqnarray}
The first relation suggests an energy dependent Planck's constant,
a phenomenon we have found before in non-linear realizations of
Lorentz invariance~\cite{leejoao1}.
Given that we have assumed $f=g$ the second relation
leads to trivial definitions of creations and annihilation
operators:
\begin{eqnarray}
\alpha^a_n&=&\sqrt n a^a_n\\
\alpha_{-n}^s&=&\sqrt n a^{a\dagger}_n
\end{eqnarray}
with
\begin{equation}
[a^a_n,a^{b\dagger}_m]=\delta_{mn}\eta^{ab}
\end{equation}
Should $f\neq g$ the whole quantization procedure should be
different. One may argue that the extra non-linear terms may be
seen as an interaction, and the system described by means of an
$S$ matrix. If this is true one should bear in mind that $f/g$
will still appear in the definition of creation and annihilation
operators and thus in the asymptotic states.

\section{The quantum constraints and the spectrum}\label{quantum}

We now examine the constraints and their enforcement at quantum
level, starting with the Hamiltonian constraint. Using
Eqns.~(\ref{xs}) and (\ref{ps}), the Hamiltonian (\ref{ham}) may
be written in terms of the amplitudes $\alpha^a_n$. The quantum
Hamiltonian, however, should contain only normal ordered
$\alpha^a_n$, and so an ordering constant $a$ has to be added at
quantum level. The result is:
\begin{equation}
H=\int d\sigma {\cal H}={{\bf P}_{CM}^2\over 2\pi fT} + {g\over 2}
{\left(\sum_{n\neq 0}|\alpha^a_n|^2 -a\right)}
\end{equation}
By setting $H=0$ we this arrive at the string center of mass
dispersion relations:
\begin{equation}\label{cmdisp}
{\eta_{ab}{\bf P}^a_{CM}{\bf P}^b_{CM}\over fg} =-M^2=2\pi
T{\left(a-\sum_{n\neq 0}|\alpha^a_n|^2 \right)}
\end{equation}
 
Given (\ref{alphcom}), it is convenient to 
define,
\be
\beta^a_n={\left(g\over f\right)}^{1/4}\alpha^a_n
\ee
so that their algebra is energy independent. It is also convenient to
define,
\be
\beta^{a}_{0}= { 1 \over f^{3/4}g^{1/4}} {\bf P}^{a}_{cm}
\ee
Their algebra is
\begin{eqnarray}
{\left[{\bf x}^a,{\beta}^{b}_{0}\right]}&=&
i \left ( { f \over g} \right )^{1/4} \delta^a_b\\
{\left[\beta_n^a,\beta_{m}^b\right]}&=& n\eta^{ab}\delta_{n+m}
\label{betacom}
\end{eqnarray}
We then define, as usual,
\be
{\tilde M}^2 = \sum_{m>0} \beta_{-m}\cdot \beta_m
\ee
In terms of these, the modified generators of the Virasoro algebra are
\begin{equation}
L_0= {1 \over 2\pi T f} [{\bf P}^2_{cm} + \sqrt{fg} {\tilde M}^2 ]
\end{equation}
\begin{equation}
L_l= \sqrt{fg}  \sum_{m>0} \beta_{l-m}\cdot \beta_m
= \sqrt{fg}  {\tilde L}_l
\end{equation}
where ${\tilde L}_l$ are, for $l \neq 0$,
the conventional Viraroso generators.
The algebra is
\begin{equation}
[ L_m , L_n ] = \sqrt{fg} (m-n) L_{m-n} + fg {D m(m^2-1) \over 12}
\delta_{m+n} . 
\end{equation}
We see that the anomaly appears to be  energy dependent,
\begin{equation} c=fg
{D m(m^2-m) \over 12}
\end{equation}

Of course, as long as $fg>0$ for all ${\bf P}_0$ we can use the rescaled
generators $\tilde{L}_n = (fg)^{-1/2}L_n$, which have the conventional Virasoro algebra, with energy independent anomaly.   However, as we will see shortly, there are interesting cases in which $fg$ vanishes for finite ${\bf P}_0$. In these cases we should be careful about
which version of the Virasoro algebra is defined  on the whole space of physical states. 

Now we study the spectrum of the theory. It is most convenient to
define $k$ as \be {\bf P}^a_{cm}|0,k> = k^{a}|0,k> \ee as ${\bf
P}^a_{cm}$ is in fact by (\ref{xs}) and (\ref{ps}) the quantity
that defines the velocity of the center of mass of the string.

The ground state is defined as usual by $L_n |0,k> =0$. The
ground state energy is given by
\begin{equation}
{\bf P}^2_{cm}= k^{2} = 2\pi T a f \sqrt{fg}
\label{ground}
\end{equation}
where $a$ is the usual energy independent constant resulting from
the normal ordering of ${\tilde M}$.

We see from these relations that the spectrum of the deformed
string will be the same as the spectrum for the ordinary bosonic
string, so long as we express momentum in terms of the non-linear
variable \f \tilde{p}^{a}= f^{-3/4}g^{-1/4}{\bf P}^a_{cm} \ff
If  $ f \sqrt{fg}>0$ and is non-vanishing and non-singular, all
the standard results on the string spectrum will go through
regarding the elimination of ghosts, the extistence of a tachyon,
$a=1$, etc.  For example, we see directly from (\ref{ground}) that
so long as $ f \sqrt{fg}>0$ the ground state remains tachyonic.
However, if we choose to violate this condition, we can eliminate
the tachyonic ground state, as we will describe below. With a suitable
choice of functions this can be done without reintroducing ghosts
into the theory.

\section{Examples}\label{examples}

\subsection{A simple example}

Let us finally look at several examples that illustrate how much
freedom is allowed by string theory, with regard to deforming the
energy-momentum relations.  We have found that to keep the
embedding degrees of freedom linear, we must take $f=g$. We have
also found that the character of the spectrum is unchanged if
$f^{2}$ is positive, non-vanishing and non-singular.

One consequence is that a varying speed of light is ruled out
~\cite{mof1,am,nc,ncinfl}.
By computing $c=dE/dp$, for $M=0$, one finds that massless modes always move
at the speed of light. On the other hand it is easy to accommodate the
dispersion relation discussed in~\cite{leejoao}, by  choosing
\be\label{f1}
f=g=1-L_P {\bf P}^{CM}_{0}.
\ee
Such a dispersion relation leads to a modified mass-energy
relation~\cite{leejoao}. Setting $l^2=1/(2\pi T)$ (the string ``length''),
the usual string spectrum mass spectrum (inferred from the right hand
side of~(\ref{cmdisp})) is
\be\label{mspec}
l^2M_n^2=N-1\, .
\ee
where $N= \sum_{n\neq 0}|\alpha^a_n|^2$.
However, the rest energy
spectrum is now
\begin{equation}
E_n={M_n\over 1+ M_nL_P}
\end{equation}
If $l \gg L_P$, the lowest string states are
uncorrected, but as the string energy approaches the Planck
energy, states accumulate just below $E_P=L_P^{-1}$ and can never exceed
this energy. This is precisely the property sought in~\cite{leejoao},
ensuring an invariant border between classical and quantum gravity.

\subsection{Eliminating the tachyon}

It is also possible to choose deformations for which the ground
state with negative $M^2$ is NOT a tachyon, a phenomenon already
discussed in the context of neutrino flavour states~\cite{neut}.
As we are about to see, this can be done by choosing deformations such
that $f^{2}$ is not positive over its entire range. Note that
there is nothing wrong with a negative $f^2$, because even though
$f$ is then imaginary, it always leads to real Lorentz
transformations (see~\cite{leejoao} for how to construct them).

We can for example choose \be \label{f2best} f^2({\bf
P}^{CM}_0)=1-(L_P{\bf P}^{CM}_0)^2 \ee This fails to be positive
only for high energies ${\bf P}_{CM}^0 > L_{P}^{-1}$. We get the
rest mass-energy relation: \be ({\bf P}_{CM}^0)^{2}={M^2\over 1+
(L_P M)^2} \ee Using (\ref{mspec}) leads to the energy spectrum.
In this case, if we choose the string scale $l<L_P$  the tachyon
state ($N=0$) is no longer a tachyon. From (\ref{ground}), we have
that at rest, \be ({\bf P}_{CM}^0)^{2} = {2 \pi T \over {L^{2}_{P}
\over l^{2}}-1 } \ee Thus, whenever $L_{P}> l$ we have $({\bf
P}_{CM}^0)^{2}>0$. Indeed the $N=0$ state is the only state with
${\bf P}_{CM}^0> L_{P}^{-1}$; states with $N\gg 1$ accumulate just
under $L_{P}^{-1}$. The lowest energy state is the first ``excited''
state $N=1$ for which $M^2=0$. As a result, the rest of the
spectrum is as in the ordinary bosonic string, expressed in terms
of $\tilde{p}^{2}= f^{-2}{\bf P}^2_{cm}$.

Another example is given by \be f^2({\bf P}^{CM}_0)={(\lambda {\bf
P}^{CM}_0)^2\over 2}\left( \pm \sqrt{1+\left( 2\over \lambda {\bf
P}^{CM}_0\right)^2} -1\right) \label{f2} \ee where $\lambda$ is a
length scale. The negative branch is chosen whenever $({\bf
P}^{CM}_0)^2<0$. With this choice we can use $E^2/f^2=M^2$ to find
the rest mass-energy relation: \be\label{emc2}
E^2_n={M^2_n\over 1+{1\over (\lambda M_n)^2}} \ee If $l>\lambda$
(a condition satisfied if $\lambda=L_P\ll l$), then the $N=0$
state has $M^2=-1/l^2$, but its rest energy squared is positive.
The ground state is again the massless state $N=1$ (for which both
$M$ and the rest $E$ are zero).  The $N=0$ state has negative
$M^2$ but behaves like a normal massive particle.

% (this phenomenon
%was previously found in the context of neutrino flavour
%states~\cite{neut}).

\section{Conclusions}

In this paper we introduced deformed dispersion relations into
string theory, considering the bosonic string, and using
perturbative canonical quantization methods.
The results found here are only a first step, and more work is
required to see if a consistent theory can be constructed along these
lines.

The basic results we found may be summarized as follows.

\begin{itemize}

\item{}The deformations of the string action
affect both the vibrational and center of mass modes.
In most cases we find that the vibrational modes no longer
decouple, and that only a very specific class of deformations
 preserve mode decoupling, those with $f=g$. This
 condition implies that
 the speed of light remains energy independent.  Thus, it appears
 that the observation of an energy dependent speed of light would be
 difficult to fit into a consistent string theory.

\item{}So long as $fg>0$ on the whole space of states, we can define
conventional Virasoro generators, and the spectrum is as usual, but 
with a transformed center of mass energy and momentum. But there
are intersesting cases in which $fg$ vanishes at finite ${\bf P}_0$, in which
case care must be used to define the Virasoro generators for all states. 
The result can be an energy dependent central charge. It may still 
be possible to choose the ghost
action so as to cancel the energy dependence of the central charge;
this remains an interesting question for future work.

\item{}For a large class of theories , those for which $f=g$ and  $f^{2}>0$
and  non-singular for its whole range, the usual
conclusions concerning the elimination of ghosts and the presence
of a tachyon hold.

\item{}By choosing $f^{2}<0$ for energies near the string scale,
the tachyonic mode of the bosonic string can be eliminated.

\item{} The choice
\be
f^2=1-(L_P{\bf P}^{CM}_0)^2
\ee
with $l_{P} < l$
is promising. It appears to produce a bosonic string without
a tachyon, and it also appears to be ghost free.

\end{itemize}

\section*{Acknowledgements} We would like to thank A. Tseytlin
and J, Schwarz  for
helpful comments. JM thanks the Perimeter Institute for support
and hospitality while this work was being done. LS was supported
during the initial phase of this work
by a gift from the Jesse Phillips Foundation.

\section*{Appendix: Particle analogue}

In order to understand the significance of the $f=g$ case, we now
consider a simple model of a relativistic massive particle in $D$
dimensional Minkowski spacetime moving in a static potential. The
action is given by
\begin{equation}
S=\int dt p_a \dot{x}^a - N {\cal H}
\end{equation}
where the Hamiltonian constraint is
\begin{equation}
{\cal H} = {f(E) \over 2} \eta^{ab}p_a p_b + g(E) [ m^2 + V[x^i ]]
\end{equation}
where $f$ and $g$ are as before functions of $E=p_0$ and
the potential is a function only of the spatial coordinates $x^i$, so
that $E$ is still a conserved quantity.
We find that the equations of motion are,
\begin{equation}
\ddot{x}^i = -{fgN \over m} {\partial V \over \partial x^i}
\end{equation}
\begin{equation}
\ddot{x}^0 = -{N^2 \over m} p_i {\partial V \over \partial x^i}
[fg^\prime -f^\prime g]
\end{equation}
So that we see that there is an anomalous acceleration of $x^0$
unless we eliminate the last factor by choosing $f=g$.  In this
case we note that if we do not make such a choice we will not be
able to fix the gauge in which $x^0 = C t$, which appears to
contradict reparameterization invariance. The resolution of this
apparent paradox is that when there is a potential and these
modified energy momentum relations we cannot assume that a
relativistic particle does not reverse direction in time, because
there will be trajectories that pass through $\dot{x}^0=0$.  So
the failure of the $x^0$ equations of motion to linearize appears
necessary if these trajectories are to be included as part of the
theory.


\begin{thebibliography}{99}
\bibitem{review}P. Biermann and G. Sigl, Lect. Notes Phys. 576 (2001) 1-26.
\bibitem{crexp}M. Takeda et al,  Astrophys.J. 522 (1999) 225;
Phys.Rev.Lett. 81 (1998) 1163-1166.
\bibitem{amel}G. Amelino-Camelia et al, Int.J.Mod.Phys. A12, 607-624, 1997;
G. Amelino-Camelia et al, Nature 393:763-765, 1998.
\bibitem{amel1}J. Ellis et al, Astrophys.J.535, 139-151, 2000.
\bibitem{liouv}J. Ellis, N.E. Mavromatos and D. Nanopoulos,
Phys. Rev. D63, 124025, 2001; ibidem astro-ph/0108295.
\bibitem{cosmicray}G. Amelino-Camelia and T. Piran,
Phys.Rev. D64 (2001) 036005.
\bibitem{amelstat}G. Amelino-Camelia, Int. J. Mod. Phys. D11, 35, 2002;
gr-qc/0012051; G. Amelino-Camelia, Phys. Lett. B510, 255-263, 2001.
\bibitem{gli}J. Kowalski-Glikman, Phys. Lett. A286, 391-394, 2001
\bibitem{leejoao1}J. Magueijo and L. Smolin, Phys. Rev. D67, 044017, 2003.
(HEP-TH 0102098); N.R. Bruno, G. Amelino-Camelia, J. Kowalski-Glikman,
Phys. Lett. B522, 133-138, 2001.
\bibitem{leejoao}J. Magueijo and L. Smolin,
Phys.Rev.Lett. 88 (190403) 2002.
\bibitem{loopdef}L. Smolin, hep-th/0209079.
\bibitem{loopdef1}G. Amelino-Camelia, L. Smolin, A. Starodubtsev,
hep-th/0305055.
\bibitem{2+1} L. Freidel, J. Kowalski-Glikman, L. Smolin,
{\it  2+1 gravity and Doubly Special Relativity},
hep-th/0307085.
\bibitem{rovelli}C. Rovelli,  Living Rev. Rel. 1 (1998) 1.
\bibitem{carlip} S. Carlip,  Rept.Prog.Phys. 64 (2001) 885.
\bibitem{pol}J. Polchinski, hep-th/9611050.
\bibitem{GSW}M. B. Green, J. H. Schwarz and E. Witten,
Superstring Theory." Cambridge University
Press ( 1987).
\bibitem{stringref}S. Forste, hep-th/0110055.
\bibitem{mof1}  J. Moffat, Int. J. of Physics D {\bf 2}, 351 (1993); J.
Moffat, Foundations of Physics, {\bf 23}, 411 (1993) .
\bibitem{am}  A. Albrecht and J. Magueijo, Phys. Rev. D {\bf 59}, 043516
(1999).
\bibitem{nc}S. Alexander and J. Magueijo, hep-th/0104093.
\bibitem{ncinfl}S. Alexander. R. Brandenberger and J. Magueijo,
Phys.Rev. D67, 081301, 2003.
\bibitem{neut}M. Blasone, J. Magueijo and P. Pires-Pacheco,
hep-ph/0307205.



\end{thebibliography}
\end{document}